\documentstyle[12pt]{article}
\begin{document}
\begin{center}
{\Large {\bf KINETICS OF MOISTURE ABSORPTION IN MIXTURES FOR IRON FOUNDRY}}\\

\vspace{1cm}
Ernesto Villar-Coci\~na, Eduardo Valencia-Morales, R\'omulo Gonz\'alez- Rodr\'{\i}guez\\

Physics Department, Central University of Las Villas, CUBA.
\end{center}

\vspace{1cm}
\begin{abstract}

  The moisture absorption in granulated materials used in foundry technologies is analyzed. The absorption process has a diffusive behavior mainly. A simple experimental technique, in which the wet weight increment was recorded as the experimental parameter and an analytic method with computing procedure to find the parameters characterizing the process was used. The determination of these parameters by traditional methods is a very difficult task so, very refined and expensive trials are needed. The fitting of the model permits to determine the diffusion coefficient and the moisture concentration in the separation surface between the sample and the environment.  The concentration profiles are established for different times. Finally, the possibility of occurrence of superimposed diffusive processes in some materials is analyzed and the diffusion coefficient and the amount of moisture incorporated by each process are calculated.\\
\end{abstract}.\\\\\\\\


\newpage
\section{Introduction}

  It is known that some porous materials are highly hygroscopic and when put in typical wet tropical climates they incorporate moisture and this influences their mechanical properties.  \\
  An example of this, is the negative influence of moisture in granulated materials based on silica sand-sodium silicate and silica sand-sugary aggregates (molasses), both used in foundry. This negative influence on the properties of the materials (loss of mechanical strength, superficial resistance, etc.) \cite{Valencia, Maka, VaGal}is due to their higly hygroscopic behavior. \\ 
  Devices made of these materials acquire, during long expositions in wet climates, a dangerous increment of moisture. As a result of this, during the casting process, a high gas generation take place, which produces defects on the surface of the casting pieces \cite{Valencia, VaGal}. For that reason, the kinetics of moisture absorption in the above granulated materials has scientific and technological importance.\\ 
  It is reported \cite{Valencia, Maka} that, for the same raw materials used in mixtures for foundry, the hygroscopic properties are directly related with the behavior of the agglutinating substances.\\
  Hygroscopicity, as physical phenomenon, is very complex and it can be framed within the general principles of adsorption and diffusion of a gas in a solid absorbent. The moisture adsorption process in these solids elapses practically in a short time \cite{DeBoher}. The physical phase formed can be constituted by only one layer of molecules of water or there can be a sudden condensation of the gas, forming several layers of molecules on the absorbent surface. The time in which this process ocurs can be negleted compared with the later diffusive processes \cite{DeBoher}. For that reason, an initial moisture concentration on the surface(for fixed relative humidities of the  environment) can be considered constant during the whole diffusive process.\\  
  According to this, which is rigorously verified in the experimental practice, the hygroscopicity research in the above materials is simplified to a diffusive problem considering physical diffusion of water \cite{Valencia}.\\ 
  For numerous industrial applications, simple but accurate models describing the natural phenomena are required. Such models have practical and economical importance \cite{Dincer}. Especially necessary is the development of models and methods to determine the process parameters: the diffusion coefficient, the moisture concentration on the separation surface (sample-surrounding atmosphere (interface)) and the amount of moisture incorporated to the sample.\\
  In this work  the wet weight increment is recorded experimentally, which can be easily determined. An analytical method with computing procedure allowed calculation of all the process parameters, including moisture concentration on the interface, which is not easy to obtain by traditional treatments\cite{Crank, Farad}and requires very expensive trials for their determination (e.g. microanalysis methods).\\
The diffusive model fitted by computerized methods permitted to determine the parameters, as the diffusion coefficient and the initial concentration of moisture on the interface. Thereafter, the concentration profiles at different time instants in the materials were established. For samples where the superimposed diffusive processes took place, a composed diffusive model was applied and the diffusion coefficients as well as the water amount incorporated by each process were determined.\\
  This procedure constitutes a cheap and satisfactory description of a phenomenon with scientific and technological interest and it can also be used for teaching purposes to study the kinetics of moisture absorption in these materials.

\section{Materials and Methods}

  The mixtures analysed were composed fundamentally by silica sand and sodium silicate or molasses as agglutinants. Their compositions are shown in Table 1. \\
  Both components were mixed in a roller mixer during 1.5 to 2 minutes (silica sand - sodium silicate mixtures) and 4-5 minutes (silica sand - molasses mixtures). Afterward standard cylindrical samples were made and compacted by three hits. Thus, the samples were put in porcelain capsules and exposed to heat treatment (silica sand - molasses mixtures) or blown with C0$_2$ during 60 seconds (silica sand - sodium silicate mixtures). The heating of the silica sand - molasses mixtures was performed in a drying-chamber at $210\:^oC \pm 1\: ^oC$ during 50 minutes and subsequently the samples were cooled to environment temperature in a hermetic dryer. Thereafter the initial weights of the samples were measured and they were placed in a climatic chamber at constant relative humidity (RH) and temperature \cite{Valencia, VaGal}. \\
  The wet weight was determined by using an analytical balance ($\pm 0.0001g$) at different times.\\ 

\begin{center}
\begin{tabular}{|c|c|c|c|c|c|c|}\hline\hline
No & Mixture & Silica  & Sodium & Molasses (\%)& D (mm)& L (mm) \\ 
   &         & sand (\%)&  silicate (\%)&        & (diameter)& (height)\\  \hline\hline
1 & Sand silica- & 94& 6 & & 58 & 7.5 \\ 
   & sodium silicate & & & & & \\ \hline 
2 & Sand silica- & 95 & & 5 & 50 & 5.2 \\
   & molasses& & & & & \\ \hline
3 & Sand silica- & 90 & & 10 & 50 & 5.2\\ 
   & molasses& & & & &  \\ \hline
\end{tabular}
\normalsize
\end{center}

{\bf Tabla 1}. Chemical composition of the materials and dimensions of the capsules. \\

\section{Formulation of the problem }

  Moisture absorption is simplified to a diffusive problem considering the physical diffusion of water, therefore it is necessary to solve the continuity equation or second Fick's law, in non-stationary states with the initial and boundary conditions imposed by the given situation. Samples, permeables only for a face, were finite round sheets (pastilles) with zero  environmental moisture as initial condition. It simplifies the mathematical problem to determine the diffusion coefficient which not depends on the geometry and boundary conditions. This effective diffusion coefficient (D) is more likely governed by the nature of the substance, the particle sizes, and by the way how these are compacted.\\ 
  By choosing this geometry, the three-dimensional diffusive problem is reduced to a unidirectional treatment (Fig.1), where D is assumed constant for each test at a fixed relative humidity. The mathematical problem is given by the equation:

\begin{equation}
\frac{\partial C}{\partial t}=D\frac{\partial ^2 C}{\partial x^2}
\end{equation}
with initial and boundary conditions given by:  \\

\begin {eqnarray}
C(x,0) &=& 0   \:\: para \: x   \in ( 0, L )\\
C(0,t) &=& Co   \:\: para \:     t > 0\\
\frac{\partial C}{\partial x}&=&0\:\: para \:x=L
\end {eqnarray}   
        
  The first boundary condition (eq.3) shows the invariance of moisture concentration (Co) in the surface and is accepted for both mixtures (silica sand-molasses and silica sand-sodium silicate), althought the silica sand-sodium silicate mixture has some amount of water (not environmental moisture) held in the gel structure initially, which is determined by the modulus of sodium silicate. This moisture contained in above mixtures does not affect the moisture absorption process qualitatively, but it changes the gradient of water concentration on the interface between the sample and the surrounding atmosphere, which could change the Co values for the same outdoors moisture conditions compared with the sand-molasses mixtures. The second boundary condition (eq.4) reflects the impermeability of the wall of the capsule in x = L. \\                       
 In order to solve the boundary problem else the Fourier method \cite{Jost, Carslaw, Tijonov}or else the method of Laplace Transform \cite{Crank} can be used. The result is:                    \\     

\begin{equation}
C(x,t)=Co \{1-\frac{4}{\pi}\:\sum\frac{1}{2n+1}\:exp(-\frac{(2n+1)^2\:\pi^2\:Dt}{4 L^2}\:\:\sin\frac{(2n+1)\pi x}{2 L}\}
\end{equation}

  This solution expresses the moisture concentration profiles according to the (x) coordinate and the (t) time in all the section of the pastille. In the experimental practice it is easier to work without measuring the concentrations but the amount of moisture that is incorporated into the sample. This amount of absorbed moisture through the A area of the permeable surface is: 
 
\begin{center}
\begin{eqnarray}
Mt&=&\int\int\int_V C(x,t)\:dxdydz\\
&=&ACoL\{1-\frac{8}{\pi^2}\:\sum_{n=0}^ \infty\frac{1}{(2n+1)^2}\:exp(-\frac{(2n+1)^2\:\pi^2\:Dt}{4 L^2}\}
\end{eqnarray}
\end{center}
where $ACoL=M_\infty$
is the amount of absorbed moisture incorporated to the sample in a sufficiently long time $t\rightarrow\:\infty$ .\\
  Traditionally, the analysis of the diffusive behavior of water into some materials is treated in terms of the relative gain of wet weight, obtaining the experimental curves of  $Mt/M_\infty\:  vs\:    t$ \cite{Crank, Farad}. This procedure gives the possibility of determining the D diffusion coefficient without knowing Co, therefore it is easier this way. However, working directly with the data of Mt vs t it is possible to determine (through the adjustment of the model (eq.7)), the Co moisture concentration on the interface and to get the real concentration profiles, which is very important.\\
\section{Results and Discussion}
  In figures 2, 3, 6 and 7 there are shown the amounts of the environmental moisture incorporated into the sample M t versus time for the silica sand - sodium silicate and silica sand-10\% molasses mixtures, which where exposed respectively to environments of relative humidities of 80 and 90\%. The curves of the fitted model are represented by solid lines. The fitting of the model permited to determine D and Co in each case. The values of the D diffusion coefficients and the Co moisture concentration on the interface are given in Table 2. In the figures the correlation coefficients r are shown.\\
  In these figures, Mt saturation values ( $M\infty$  )are larger for samples exposed under 90\% RH than for the same sample at 80\% RH, something expected due to the fact that the greater the water concentration of the environment, the greater the Co will be in the interface (Table 2), and therefore the gain of wet weight will be greater.\\
From these figures the hygroscopic power that molasses give to the mixtures is also deduced, since the relative gains of moisture in the equilibrium are greater in the 10\% molasses mixtures than in the 5\% molasses mixtures at the same environmental conditions.\\
  In figures 4 and 5, for example, the moisture concentration profiles for silica sand and sodium silicate samples are shown.\\
  A considerable variation of the concentration with the pastille depth is appreciated in the initial instants, getting the reflection on the impermeable face at short times, until, close to saturation, the moisture concentration on the sample is practically the same and equal to the Co surface concentration. The moisture concentration in the sample can be known at any depth from the surface at different time instants through these profiles.\\
  A peculiar situation is found in the silica sand - 5\% molasses mixtures. According to the results shown in Fig. 8 and 9, it is clear that specifically in the case of these mixtures, it should be assumed that there are two independent places in the structure of the mixture which can be the cause of the superposition of two different processes with diffusion coefficients $D_1 < D_2$.\\
  In the literature \cite{Frisch, Doremus, Bokshtein, Zaes, Riekert, Quereshi, Ackley} some examples are reported where difficulties during the evaluation of the diffusive process through a simple model appear. These were overcome by considering two different places of diffusion or different phases in the material and, therefore, 2 diffusive superposed processes with different diffusion coefficients.\\
  The amount of moisture incorporated to the sample is then an additive superposition of solutions of type (7) where $(ACoL )^{(1)} = M\infty ^{(1)}$ and  $(ACoL)^{(2)} = M\infty ^{(2)}$ are the amounts of moisture incorporated to the sample by each process after a sufficiently long time. The supraindex 1 and 2 correspond to processes 1 and 2 respectively. Observe that the sum of $M\infty ^{(1)}$ and $M\infty ^{(2)}$ coincides with the corresponding experimental saturation value of the mixture.\\
  The superposition of diffusive processes could be explained phenomenologically in this way: the mixture is constituted by silica sand and the agglutinant, the last one joins the grains strongly by making a layer around them. This layer increases its thickness with the increase of the molasses content for the same grain size in the mixture. Furthermore, it is also known that molasses are highly hygroscopic \cite{Valencia}. We can presuppose then a principal diffusive way (principal process) through the molasses and another secondary diffusion way (secondary process) through the boundary between sand grains and molasses, which acts as agglutinant.\\
  The consideration of a simple diffusive process in some of these mixtures leads to results which are not as satisfactory as in the case of the composed process.\\
  The fact that in some cases both processes are more noticeable and in other cases only a simple process is noticed, is given by the magnitude of one process compared with the other one. When the magnitude of one process (secondary) is very small compared with the other one (principal) this last one overlaps the first one. In this case the principal process is almost not affected by the superposition and it seems that only one process is happening. An approximation to a simple process ($D_1 = D$)in this cases does not lead to large mistakes in the description process. Several authors have analyzed this problem in other systems \cite{Frisch, Doremus, Bokshtein, Zaes, Riekert, Quereshi, Ackley}.\\

  \begin{center}
\begin{tabular}{|c|c|c|c|}\hline\hline
 Mixture & Relative  & Diffusion coefficient D & Concentration in the interface Co\\ 
   & humidity (\%) & ($mm^2/s$) & ($g/mm^3$)     \\   \hline\hline
 Silica sand - & 80& $8,40.10^{-4}\pm 0,20.10^{-4}$ &$4,02.10^{-6}\pm 0,02.10^{-6}$ \\    
 silicate of sodium & & &  \\ \hline 
 Silica sand - & 90 &$1,04.10^{-3}\pm 0,03.10^{-3}$ &$9,11.10^{-6}\pm 0,06.10^{-6}$ \\
 silicate of sodium& & & \\ \hline
 Silica sand - & 80 &$7,7.10^{-4}\pm 0,3.10^{-4}$ & $8,45.10^{-6}\pm 0,06.10^{-6}$\\ 
 10\% molasses& & &  \\ \hline
 Silica sand - & 90 &$6,9.10^{-4}\pm 0,4.10^{-4}$ & $1,0.10^{-5}\pm 0,02.10^{-5}$\\ 
 10\% molasses& & &  \\ \hline

\end{tabular}
\normalsize
\end{center}

{\bf Tabla 2}. Diffusion coefficients and moisture concentration in the interface for different materials and relative humidities. \\

  In the analyzed case of silica sand-molasses mixtures, this phenomenon is more manifested in 5\% of molasses (least molasses), where the consideration of a simple diffusive process leads to results that are not so satisfactory as observed in Fig. 10 compared with the consideration of a compound process(see  Fig. 9).\\
  The values of the diffusion coefficients and the amount of environmental moisture incorporated to the sample by both processes are shown in Table 3. \\

 \begin{center}
\begin{tabular}{|c|c|c|c|c|c|}\hline\hline
 Mixture & Relative  & $D_{1}$ ($mm^2/s$) & $M\infty^{1}$& $D_{2}$ ($mm^2/s$)& $M\infty^{2}$\\ 
  & humidity(\%) & & & &     \\   \hline\hline
 
 Silica sand - & 80 &$1,03.10^{-3}\pm$  & $0,01892\pm$ &$6,55.10^{-3}\pm$ &$0.01328\pm$ \\ 
 5\% molasses & & $0,05.10^{-3}$ &$0,00092$ &$0,8.10^{-3}$ & $0.00096$  \\ \hline
 Silica sand - & 90 &$6,9.10^{-4}\pm$  &$0.05888\pm$ & $1,5.10^{-2}\pm$ &$0.02734\pm $\\ 
 5\% molasses& & $0,2.10^{-4}$ &$0.00067 $& $0,01.10^{-2}$& $0.00072 $\\ \hline

\end{tabular}
\normalsize
\end{center}

{\bf Table 3}. Diffusion coefficients and amount of environmental moisture incorporate to the materials.\\

  In the case of 10\% molasses mixtures, the principal process prevails and it can be described by a simple process. (Fig. 6 and 7). The values of diffusion coefficients and moisture concentration on the interface are shown in Table 2. \\
 The above is explained taking into account the fact that, reducing the molasses \%, the agglutinant layer thickness is reduced and the effective diffusion area is also reduced. This evidently causes the magnitude of the diffusive process through molasses to be smaller and so it does not overlap the secondary process. Fig. 8 and 9 give an idea of the magnitude of both processes and of the addition of processes.\\

\section{Conclusions}

The diffusive model assumed permits to describe the moisture absorption process that is evidenced in the experiment in these materials, determining the diffusion coefficient values and the moisture concentration in the interface (surrounding atmosphere - sample) in the fitting model process with enough accuracy.\\ 
 
It all permits to determine the concentration profiles C (x, t) in a different way than other treatments in which C (x, t)/Co is obtained, and therefore it shows the amount of moisture in the sample at any depths from the surface at different time instants.\\

Evidently, these results can be generalized to samples of finite thickness and with any geometry, but in this case the graph and adjustment of $Mt/M_{\infty}$ vs t are considered and the concentration profiles C/Co vs x are established for different time instants.\\

The occurrence of superimposed diffusive processes is evidenced for silica sand - 5\% molasses mixtures, and the values of the diffusion coefficients and the amount of moisture incorporated to the sample for both processes are determined accurately in the fitting process of a composed model.\\

\newpage

\begin{center}
{\bf Referencias}
\end{center}

\newpage

\begin{Large}
Figure Captions
\end{Large}

{\bf Figure 1}. Diagram of the finite pastille permeable only for a face.\\\\\\

{\bf Figure 2}. Gain curve of wet weight of silica sand-sodium silicate sample in environment of 80\% RH.\\
               $\bullet$  Experimental $\;\;--$ Model\\\\\\

{\bf Figure 3}. Gain curve of wet weight of silica sand-sodium silicate sample in environment of 90\% RH.\\
               $\bullet$  Experimental $\;\;--$ Model\\\\\\

{\bf Figure 4}. Concentration profiles of silica sand-sodium silicate sample in environment of 80\%RH at different time instants. \\\\\\

{\bf Figure 5}. Concentration profiles of silica sand-sodium silicate sample in environment of 90\%RH at different time instants. \\\\\\

{\bf Figure 6}. Gain curve of wet weight of silica sand-10\% molasses sample in environment of 80\% RH.\\
               $\bullet$  Experimental $\;\;--$ Model\\\\\\

{\bf Figure 7}. Gain curve of wet weight of silica sand-10\% molasses sample in environment of 90\% RH.\\
               $\bullet$  Experimental $\;\;--$ Model\\\\\\

{\bf Figure 8}. Gain curve of wet weight of silica sand-5\% molasses sample in environment of 80\% RH, considering 2 superimposed diffusive processes.\\
               $\bullet$  Experimental $\;\;--$ Model\\\\\\

{\bf Figure 9}. Gain curve of wet weight of silica sand-5\% molasses sample in environment of 90\% RH, considering 2 superimposed diffusive processes.\\
               $\bullet$  Experimental $\;\;--$ Model\\\\\\

{\bf Figure 10}. Gain curve of wet weight of silica sand-5\% molasses sample in environment of 90\% RH, considering a simple diffusive process.\\
               $\bullet$  Experimental $\;\;--$ Model\\\\\\

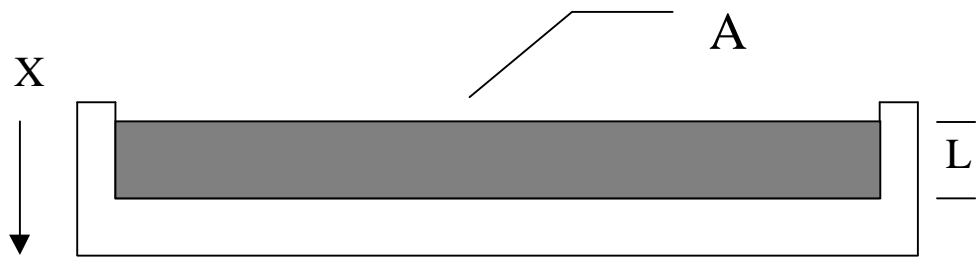
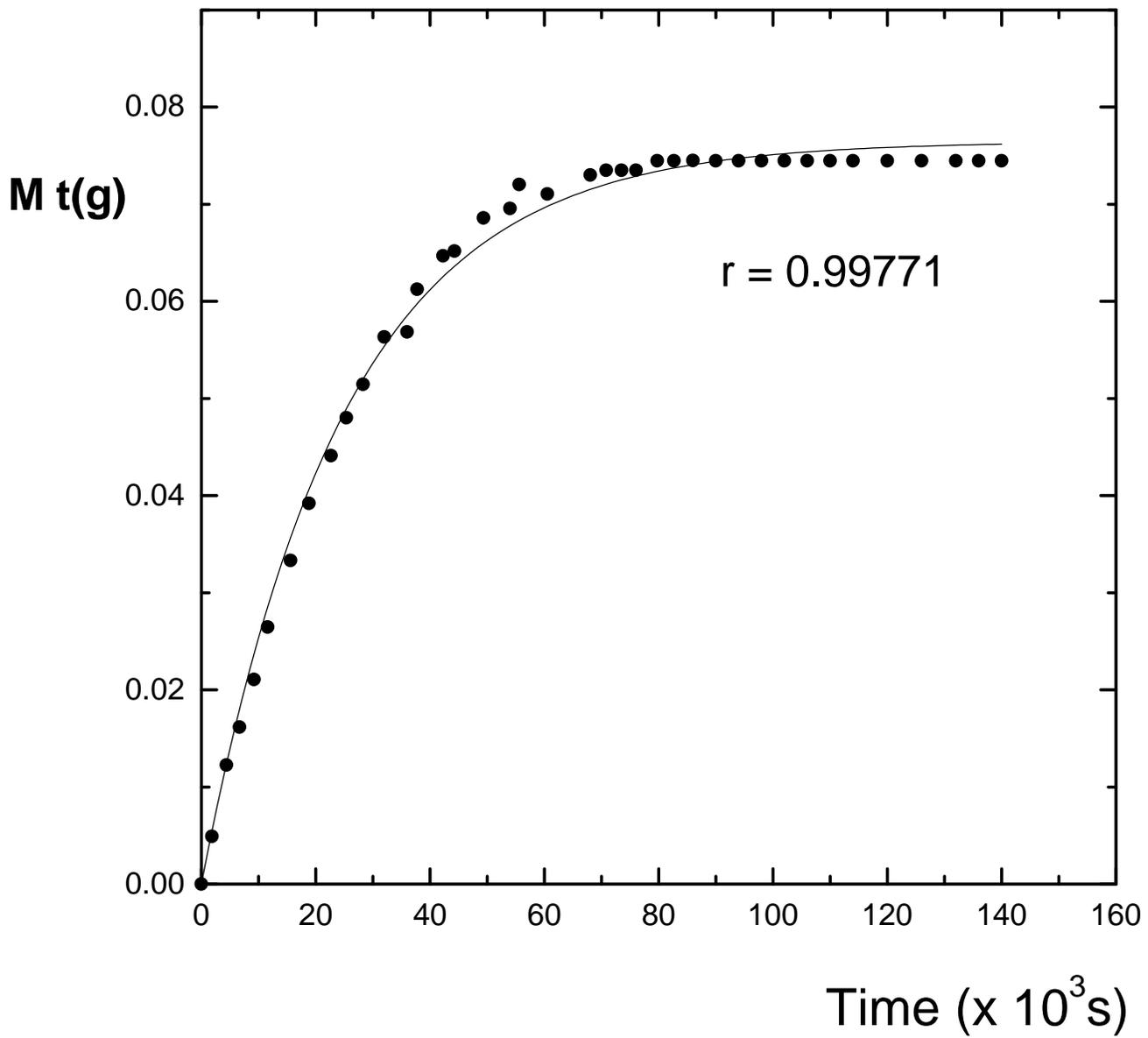
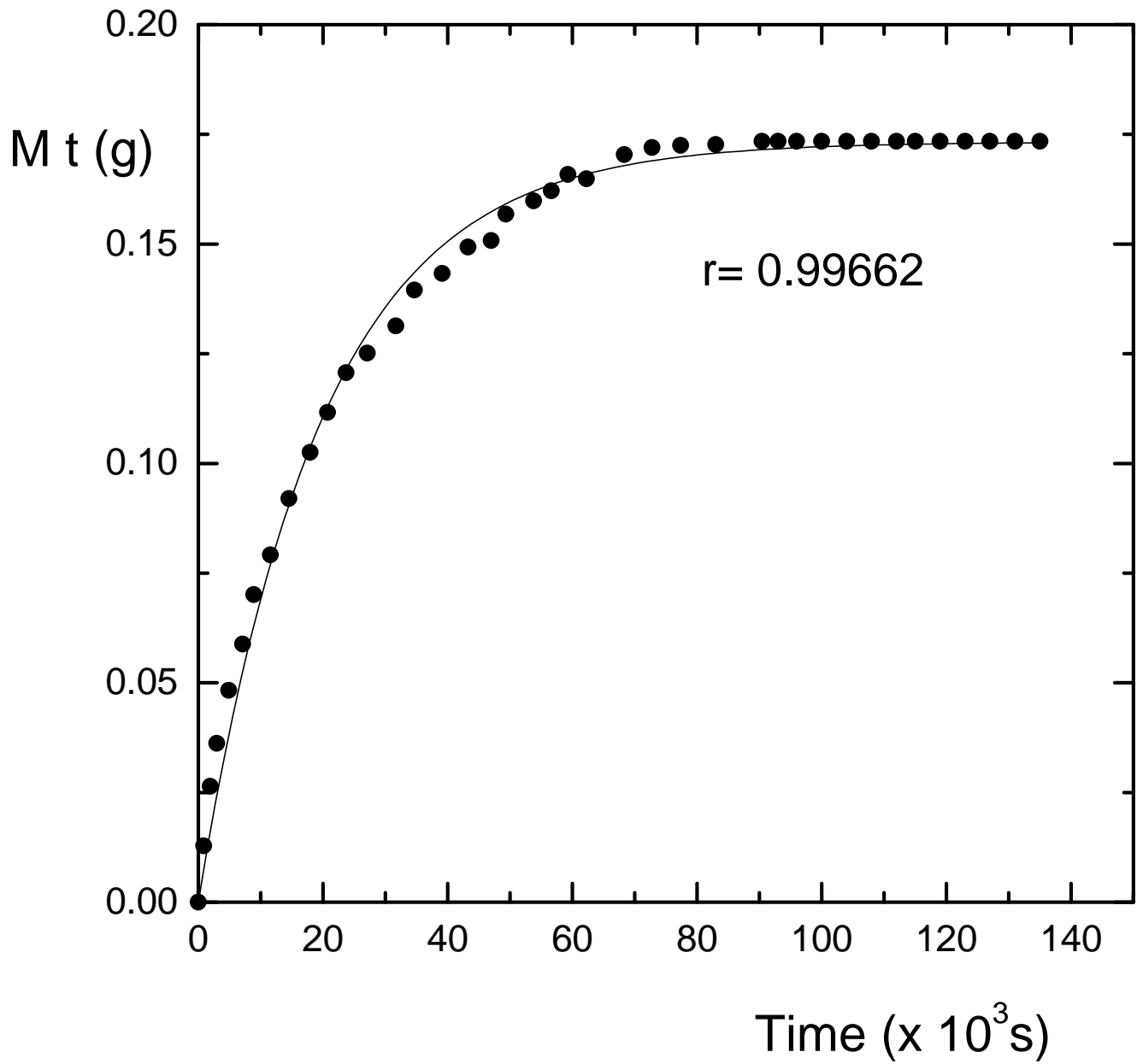
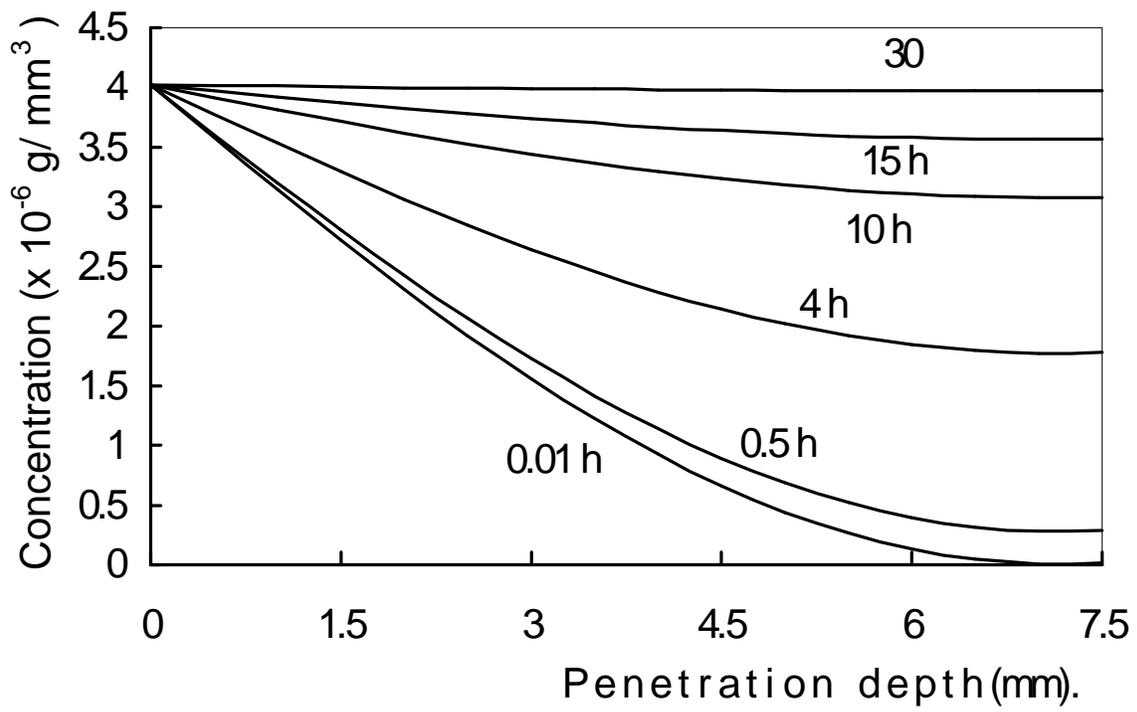
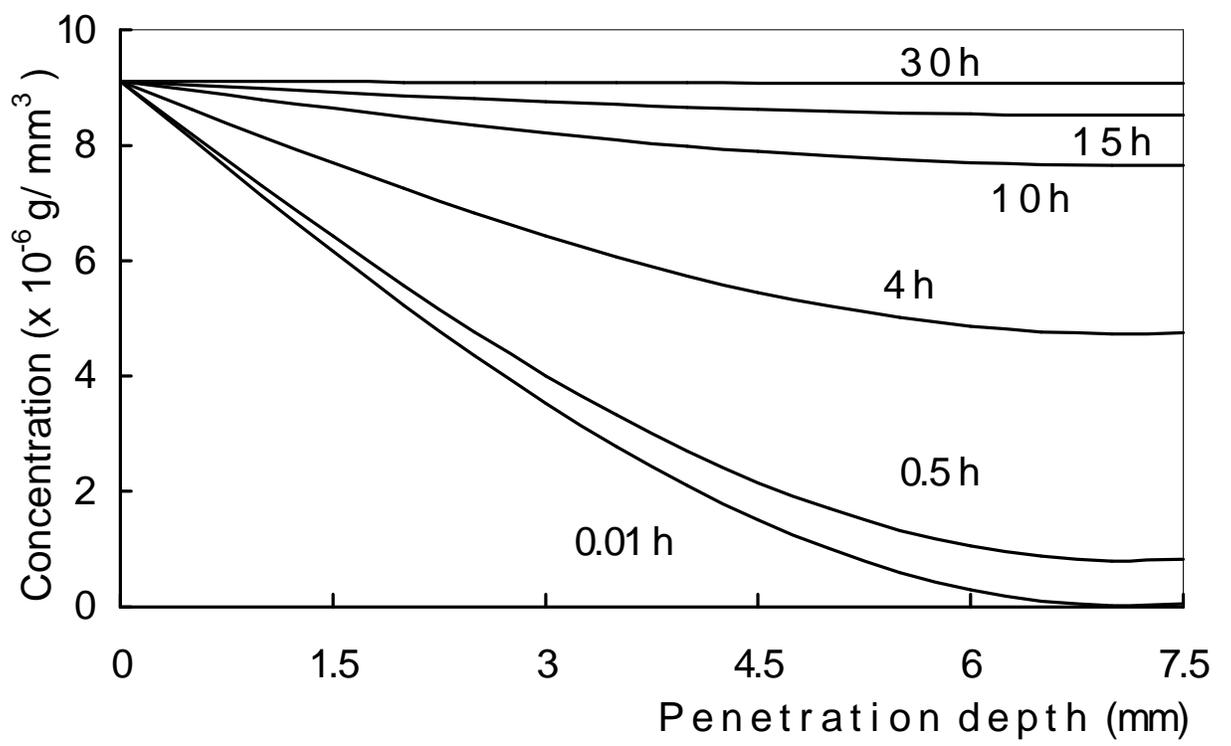
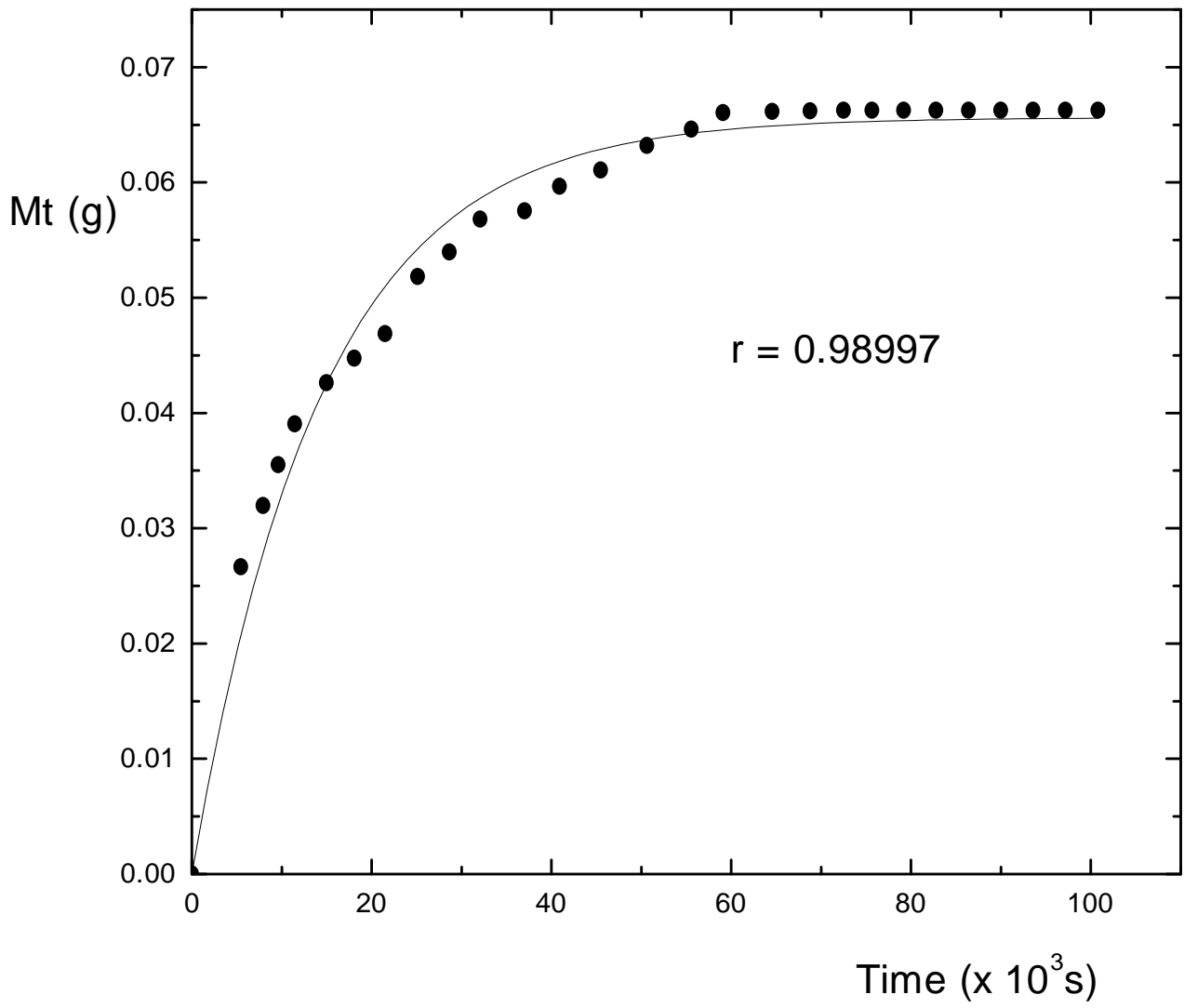
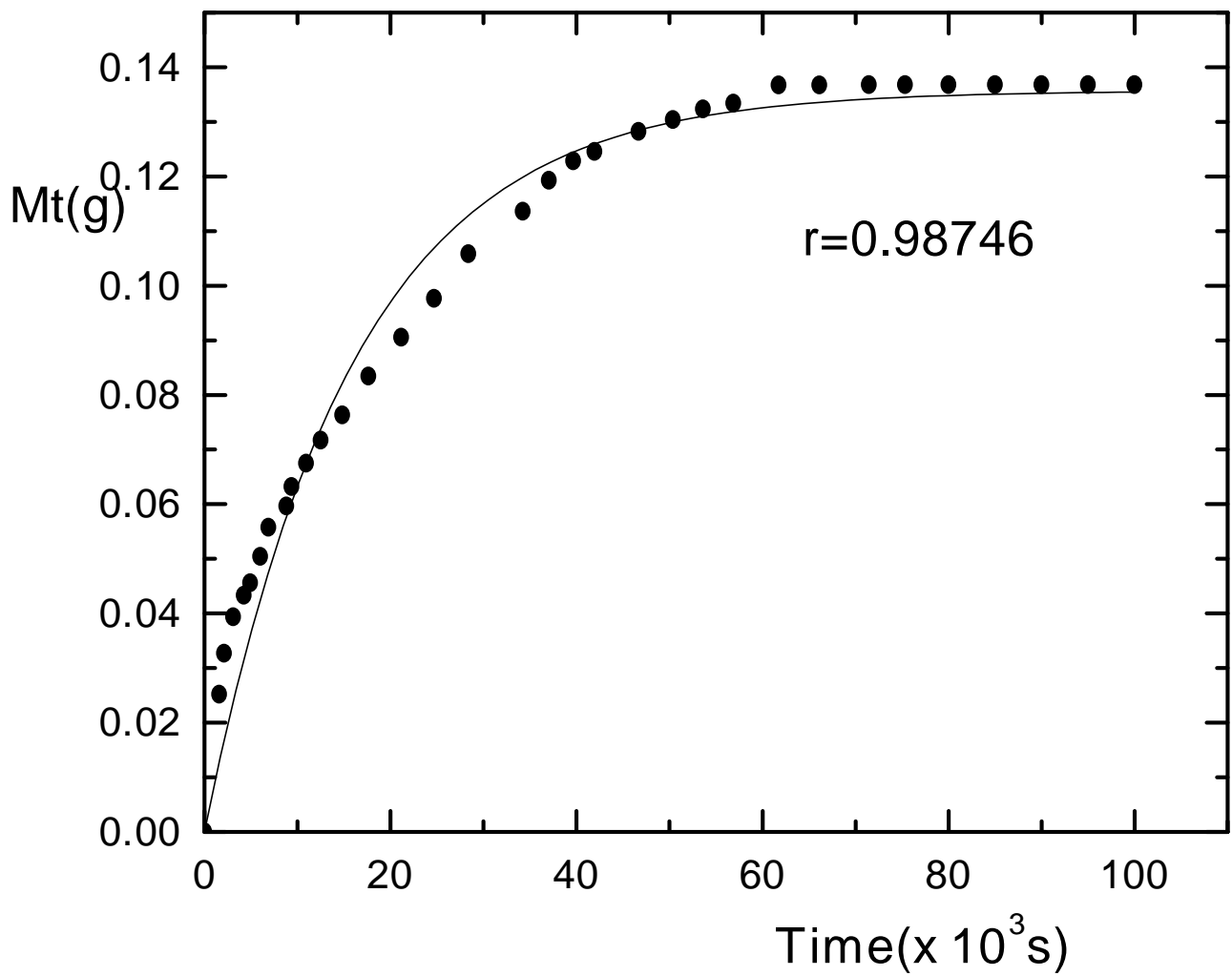
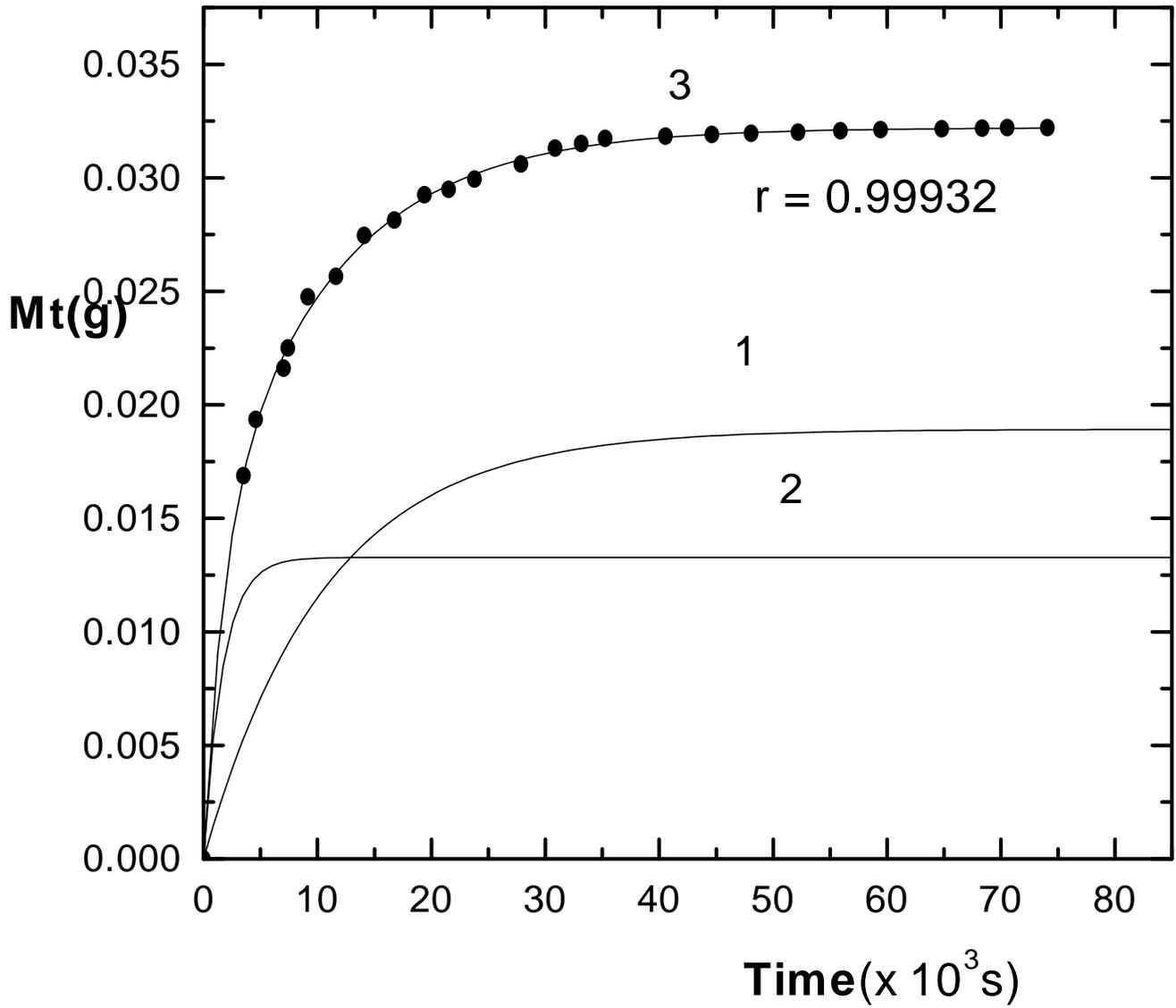
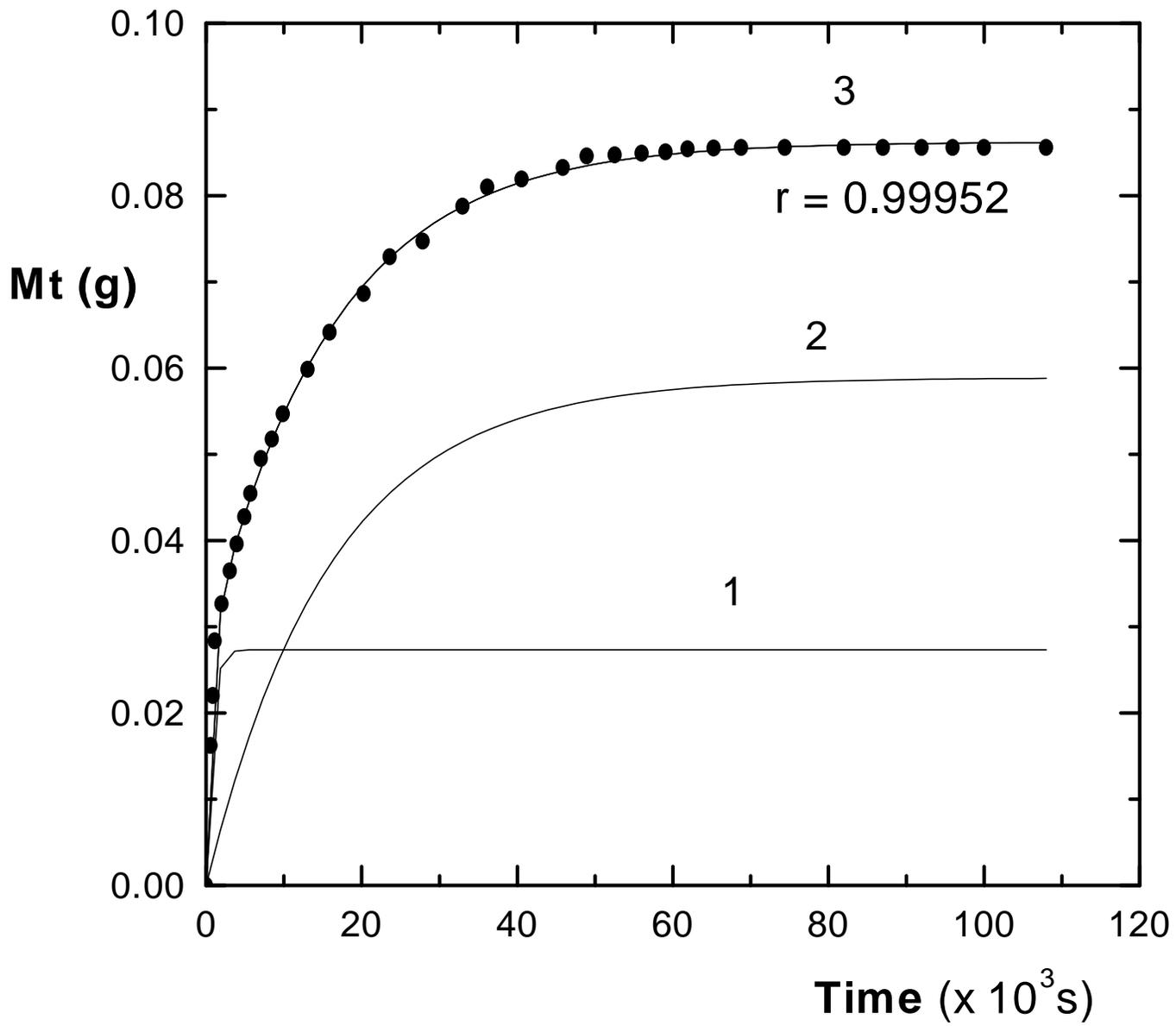
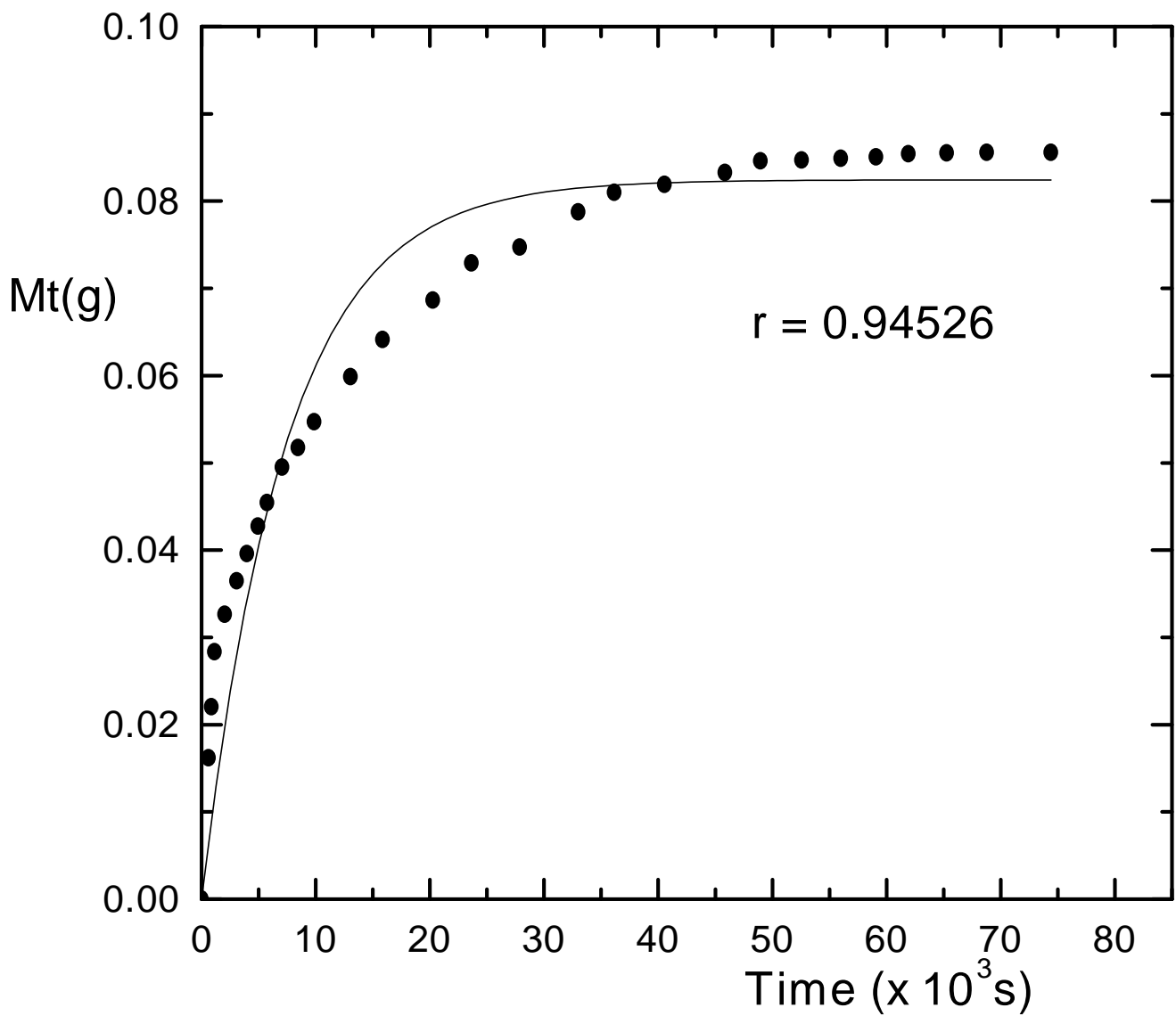
\end{document}